# Instantaneous ionization rate of $H_2^+$ in intense laser field; Interpretation of the Enhanced Ionization

**Mohsen Vafaee[1], Hassan Sabzyan[1], Zahra Vafaee[2], Ali Katanforoush[3]**

[1] Department of Chemistry, University of Isfahan, Isfahan 81746-73441, I. R. Iran

[2] Department of Mathematics, University of Isfahan, Isfahan 81746-73441, I. R. Iran

[3] School of Mathematics, Institute for Studies in Theoretical Physics and Mathematics (IPM), Tehran 19395-5746, I. R. Iran

E-mail: 1) MohsenVafaee@ sci.ui.ac.ir,  2) sabzyan@sci.ui.ac.ir

## Abstract

The fixed-nuclei full dimensional time-dependent Schrödinger equation is directly solved for $H_2^+$ in the linearly polarized laser field of $I=1.0\times10^{14}$ W/cm$^2$ and $\lambda$ = 1064 nm. Instantaneous ionization rate has been introduced and calculated by evaluating the instantaneous imaginary energy of the system. It is shown that positive (negative) values of instantaneous imaginary energy of the system represent the incoming (outgoing) instantaneous current of electron. This approach allows us to determine not only the instantaneous intensity but also the instantaneous direction of the electronic current. The transient behavior of the electron wavepacket in intense laser field can thus be probed precisely. Details of the enhanced ionization rates are studied based on the instantaneous ionization rates. This approach gives direct evidence for existence of the effect of charge-resonance-enhanced multiphoton resonances of the quasi-energy states (QES) with excited electronic states at some particular internuclear distances. Finally, Contributions of the individual time dependent Floquet QES to the overall ionization rates are evaluated.

# Introduction

Hydrogen atom and hydrogen molecular ion are two fundamental and prototypical systems which can be used to understand and extend the science of atomic and molecular physics. In intense laser field, these two apparently simple systems exhibit various complex phenomena.[1-3] An interesting and complicated effect is the enhancement of the ionization rate of $H_2^+$ as a function of inter-nuclear separation which results in maxima at some critical inter-nuclear H-H separations. Indirect[4,5] and direct[6] evidences has been reported for a similar behavior in more complex molecules. In order to characterize and interpret this enhanced ionization rates, which was observed previously at some critical inter-nuclear distances[7-12], recent theoretical studies on the $H_2^+$–laser interaction have been concentrated on the calculation of the ionization rate as a function of inter-nuclear distance R.[13-25]

For the molecules subjected to short intense laser pulses, molecular dissociation is accompanied by ionization. At high intensities ($I > 10^{14}$ W/cm$^2$) for which the ionization process is generally faster than the dissociation process, it is possible to study molecular ionization processes independently at fixed inter-nuclear separations. The ionization rate of $H_2^+$ near the equilibrium inter-nuclear separation of 2 au is relatively small and thus dissociation proceeds more easily than ionization does. The molecules in intense laser fields are strongly aligned due to the acquired torque from the laser field. Moreover, the ionization process is relatively fast compared to the vibrational motion of the nuclei. Therefore, the numerical solution of the full dimensional time-dependent Schrödinger equation (TDSE) of $H_2^+$ system in intense laser field is reduced to the numerical solution of the electronic TDSE as the fixed nuclei approximation, which is also experimentally well acceptable, can well be employed.

This paper is structured as follows. At first, the numerical method used for the modeling of the $H_2^+$ system is introduced and the details of simulations are presented. Next, results obtained for the $\lambda$ = 1064 nm and I = 1×10$^{14}$ W cm$^{-2}$ laser pulse are presented and compared with the results of other calculations reported recently. Then, the instantaneous ionization rate (IIR) is introduced and the results are displayed and discussed. Unless otherwise stated, atomic units have been used throughout.

## Numerical solution of the TDSE

In this study, ionization of the hydrogen molecular ion $H_2^+$ under intense linearly polarized pulse of laser fields is simulated by direct solution of the fixed-nuclei, full dimensional electronic TDSE equation for $\lambda = 1064$ nm wavelength and $I=1.0\times10^{14}$ W/cm$^2$ intensity. Time-dependent Schrödinger equation in the cylindrical polar coordinate system for $H_2^+$ molecular ion located in the laser field of $E(t) = E_0 f(t)\cos(\omega t)$ parallel to the $z$ axis (inter-nuclear axis) reads as

$$i\frac{\partial \psi(z,\rho,t)}{\partial t} = H(z,\rho,t)\psi(z,\rho,t), \qquad (1)$$

where the total 3D electronic Hamiltonian is given by [15,17]

$$H(z,\rho,t) = -\frac{2m_p + m_e}{4m_p m_e}\left[\frac{\partial^2}{\partial \rho^2} + \frac{1}{\rho}\frac{\partial}{\partial \rho} + \frac{\partial^2}{\partial z^2}\right] + V_C(z,\rho,t) \qquad (2)$$

in which potential $V_C(z,\rho,t)$ of the interacting system is given by

$$V_C(z,\rho,t) = \frac{-1}{\left((z\pm R/2)^2 + \rho^2\right)^{\frac{1}{2}}} + \left(\frac{2m_p + 2m_e}{2m_p + m_e}\right)zE_0 f(t)\cos(\omega t) \qquad (3)$$

with $E_0$ being the laser peak amplitude, $\omega = 2\pi\upsilon$ the angular frequency, and $f(t)$ the laser pulse envelope which is set as

$$f(t) = \begin{cases} \dfrac{1}{2}\left[1-\cos\left(\dfrac{\pi t}{\tau_1}\right)\right] & \text{for } 0 \leq t \leq \tau_1 \\ 1 & \text{for } \tau_1 \leq t \leq \tau_1 + \tau_2 \\ \dfrac{1}{2}\left[1-\cos\left(\dfrac{\pi(t-\tau_2-2\tau_1)}{\tau_1}\right)\right] & \text{for } \tau_1+\tau_2 \leq t \leq 2\tau_1+\tau_2 \end{cases}, \qquad (4)$$

with $\tau_1$ being the rising and falling time and $\tau_2$ being the duration of the laser pulse at its full-scale amplitude. In the present calculations, $\tau_1$ and $\tau_2$ are set to 5 and 15 cycles, respectively.

To solve the TDSE, Eq. (1), numerically in the cylindrical polar coordinate system, we adopt a more general nonlinear coordinate transformation $z \to \tilde{z}$ and $\rho \to \tilde{\rho}$, from which the operators in Eq. (2) become:

$$\frac{\partial}{\partial \rho} = \frac{1}{\rho'}\frac{\partial}{\partial \tilde{\rho}}, \quad \frac{\partial^2}{\partial \rho^2} = \frac{1}{\rho'}\frac{\partial}{\partial \tilde{\rho}}\left(\frac{1}{\rho'}\frac{\partial}{\partial \tilde{\rho}}\right) \text{ and } \frac{\partial^2}{\partial z^2} = \frac{1}{z'}\frac{\partial}{\partial \tilde{z}}\left(\frac{1}{z'}\frac{\partial}{\partial \tilde{z}}\right). \qquad (5)$$

An alternative set of TDSE and its norm is thus obtained;

$$\psi(z,\rho,t) = \frac{\tilde{\phi}(\tilde{z},\tilde{\rho},t)}{\sqrt{z'\rho'}}, \qquad (6)$$

$$N = \iint \rho |\psi(z,\rho,t)|^2 d\rho dz = \iint \rho |\tilde{\phi}(\tilde{z},\tilde{\rho},t)|^2 d\tilde{\rho}d\tilde{z}. \qquad (7)$$

More details of these transformations are described in our previous report [17]. By substituting Eqs. (6)-(7) to Eqs. (1)-(3), we have

$$i\frac{\partial \tilde{\phi}(\tilde{z},\tilde{\rho},t)}{\partial t} = \tilde{H}(\tilde{z},\tilde{\rho},t)\tilde{\phi}(\tilde{z},\tilde{\rho},t), \qquad (8)$$

$$\tilde{H}(\tilde{z},\tilde{\rho},t) = \tilde{H}_z(\tilde{z}) + \tilde{H}_\rho(\tilde{\rho}) + V_C(\tilde{z},\tilde{\rho},t), \qquad (9)$$

where

$$\tilde{H}_z(\tilde{z}) = -\beta \frac{\sqrt{z'}}{z'} \frac{\partial}{\partial \tilde{z}} \left( \frac{1}{z'} \frac{\partial}{\partial \tilde{z}} \right) \frac{1}{\sqrt{z'}} , \qquad (10)$$

$$\tilde{H}_\rho(\tilde{\rho}) = -\beta \left[ \frac{\sqrt{\rho'}}{\rho'} \frac{\partial}{\partial \tilde{\rho}} \left( \frac{1}{\rho'} \frac{\partial}{\partial \tilde{\rho}} \right) + \frac{\sqrt{\rho'}}{\rho \rho'} \frac{\partial}{\partial \tilde{\rho}} \right] \frac{1}{\sqrt{\rho'}} , \qquad (11)$$

$$V_C(\tilde{\rho}, \tilde{z}, t) = \frac{-1}{\left( (z \pm R/2)^2 + \rho^2 \right)^{\frac{1}{2}}} + \kappa z E_0 f(t) \cos(\omega t) . \qquad (12)$$

The differential operators in Eqs. (10) and (11) are discretized by the eleven-point difference formulae which have a tenth-order accuracy.[26] Now, for the time discretization of Eq. (8), the following propagator, derived from split-operator methods[27], has been used.

$$\begin{aligned}
\tilde{\phi}(t, t+\delta t) = & \left[ 1 + i\frac{\delta t}{4} \tilde{H}_\rho - \frac{(\delta t)^2}{48} (\tilde{H}_\rho)^2 \right]^{-1} \left[ 1 - i\frac{\delta t}{4} \tilde{H}_\rho - \frac{(\delta t)^2}{48} (\tilde{H}_\rho)^2 \right] \\
& \times \left[ 1 + i\frac{\delta t}{4} \tilde{H}_z - \frac{(\delta t)^2}{48} (\tilde{H}_z)^2 \right]^{-1} \left[ 1 - i\frac{\delta t}{4} \tilde{H}_z - \frac{(\delta t)^2}{48} (\tilde{H}_z)^2 \right] \exp(i \delta t V_C) \\
& \times \left[ 1 + i\frac{\delta t}{4} \tilde{H}_\rho - \frac{(\delta t)^2}{48} (\tilde{H}_\rho)^2 \right]^{-1} \left[ 1 - i\frac{\delta t}{4} \tilde{H}_\rho - \frac{(\delta t)^2}{48} (\tilde{H}_\rho)^2 \right] \\
& \times \left[ 1 + i\frac{\delta t}{4} \tilde{H}_z - \frac{(\delta t)^2}{48} (\tilde{H}_z)^2 \right]^{-1} \left[ 1 - i\frac{\delta t}{4} \tilde{H}_z - \frac{(\delta t)^2}{48} (\tilde{H}_z)^2 \right] \tilde{\phi}(t).
\end{aligned} \qquad (13)$$

This propagator is unitary and can be obtained by combining the classical split operator and the Crank-Nicholson method.[28]

## Time-Dependent Ionization Rates; Results and Discussion

We performed the simulation for a huge box, i.e. with (640,170) size for ($z$, $\rho$) directions. The absorber regions are set at $-320 \leq z \leq -300$ and $300 \leq z \leq 320$ for the $z$ direction and at $150 \leq \rho \leq 170$ for the $\rho$ direction. Energy of the $H_2^+$ system in the laser field can be calculated using:

$$E(t) = \int_{\rho=0}^{\rho=150} \int_{z=-300}^{z=300} \psi(z,\rho,t) H(z,\rho,t) \psi^*(z,\rho,t) \rho d\rho dz, \tag{14}$$

The instantaneous energy thus obtained can be decomposed as $E(t) = E_{Re}(t) + iE_{Im}(t)$, in which $E_{Re}(t)$ and $E_{Im}(t)$ are the real and imaginary parts of the instantaneous total energy of the system, respectively.

The imaginary energy was used previously to calculate the ionization rate in the wavepacket method [19] via the $\Gamma = -2E_{Im}$ relation. This relation has so far been used to calculate ionization rate in the Floquet method, both in the time-independent[13] and time-dependent[18] approaches. Here, we have used the imaginary part of the instantaneous energy for introducing and calculating instantaneous ionization rate, $\Gamma(t)$, as

$$\Gamma(t) = \frac{-2E_{Im}(t)}{N(t)} \tag{15}$$

in which $N(t)$ is the instantaneous norm of the wavefunction at time $t$.

Instantaneous imaginary energies of $H_2^+$ at some fixed inter-nuclear separations have been calculated by direct solution of the time-dependent Schrödinger equation in the presence of the linearly polarized laser field with I=1.0×10$^{14}$ W/cm$^2$ and $\lambda$ = 1064 nm. The overall ionization rate of $H_2^+$ at each inter-nuclear distance is calculated by averaging the calculated instantaneous ionization rates. The ionization rates are averaged from $\tau = 10$ to $\tau = 20$ cycles of the laser field. Figure 1 shows the R-dependent ionization rates calculated in the present study (●) for the $H_2^+$ system. In the same Figure, ionization rates calculated by Lu and Bandrauk (Δ)[15], Peng et al. (□)[16] and Chu and Chu (×)[18] are also given for comparison. The isolated single point on the right vertical axis (at R=14) corresponds to the ionization rate of an isolated H atom in the same laser field and is given for comparison. As can be seen from Figure 1, the four sets of calculated ionization rates are in good validating agreement, especially in the range of R>9.6. It can be seen also that the first baseline peak (spanning over inter-nuclear distances between R=4 and R=7.5) is structured. Our results show furthermore a weak peak at R=4.8 which is not predicted by previous works, and another stronger peak at R=6 which is consistent with results reported by Lu and Bandrauk[15] and Chu and Chu.[18] The strongest peak is located at about R=9.4.

It is obvious that the instantaneous ionization rate, $\Gamma(t)$, can be obtained by calculating the time-dependent norm of the wavefunction ($N(t) = \|\Psi(t)\|^2$) using

$$\Gamma(t) = \frac{-d\ln(N(t)/N(0))}{dt} \tag{16}$$

where $N(0) = \|\Psi(0)\|^2$ is the norm of the system at $t=0$. Both equations, Eq. (15) and Eq. (16), give the same results.

Evaluation of the instantaneous ionization rate (via either equations) has a distinct advantage; allowing us to follow the time-dependent ionization processes of a system and to study details of the ionization mechanism. For this purpose, the time-dependent ionization rates of $H_2^+$ in the linearly polarized laser field of I= $1.0\times10^{14}$ W/cm$^2$ intensity and $\lambda = 1064$ nm wavelength for a number of selected inter-nuclear separations have been calculated and presented in Figure 2. The R values in Figure 2 correspond to the peaks of the ionization rates presented in Figure 1.

All of the calculated time-dependent ionization rate curves shown in Figure 2 consist of repeating non-symmetric baseline peaks corresponding to one half cycle of the laser field. Each one of these baseline peaks has several sharp peaks in its structure. The relative amplitudes of these sharp peaks may change, but their relative positions remain almost constant from one baseline peak to the other. The baseline peaks, which are related to the outgoing electron current from the boundary of the system, have a delay time with respect to the peaks of the laser field. For example, our calculation shows that the delay time for R = 9.6 is about 0.85 cycle. The delay time depends on the size of the system; the larger the grid size of the system, the longer the delay time.

The mechanism giving rise to the critical distances in which ionization rates are enhanced and details of the R-dependent ionization rates have received considerable attentions from the earliest studies.[4-23] Zuo and Bandrauk[8] suggested that *charge-resonance* and positions of the two lowest levels of the $H_2^+$ system in the laser field (at its maximum amplitude) can explain the enhanced ionization peaks. Mulyukov et al.[14] believe that *over barrier ionization* out of the higher well for the most prominent peak in the ionization rate is impeded by backscattering of the electron from the hump between the wells. They argued that at certain values of R, the electron can escape more easily by undergoing a resonant transition to an energetically nearby highly excited state localized in the lower well so that the hump between the wells does not impede over barrier ionization.

Chu and Chu tried to resolve detailed mechanism of the ionization enhancement phenomenon by calculating ionization rates of the quasi-energy states (QESs).[18] They applied an ac Floquet calculation to study characteristics and dynamical behavior of complex QESs at different R. They introduced a complex-scaling generalized pseudospectral (CSGPS) technique for the determination of the complex QESs for two-center diatomic molecular systems. Chu and Chu's calculations showed that the dominant electron population remains in the $1\sigma_g$ and $1\sigma_u$ states in the laser field of $1\times10^{14}$ W/cm$^2$ intensity. Thus, the two distinctly different groups of QESs, i.e. the lower and upper groups, whose major components are the field-free $1\sigma_g$ and $1\sigma_u$, can be derived.

Because of the Floquet symmetry, all of the QESs in the lower (or in the upper) group separated by $2m\omega$ (m is integer) in energy, are in fact physically indistinguishable and contain the same information regarding multiphoton dynamics. Thus, all QESs in the lower (or in the upper) group have identical imaginary energies. They used the dynamical properties of these two sets of QESs to explore the mechanism(s) responsible for the enhanced ionization phenomenon. Their results reported also in Figure 3 indicate that the QESs in the lower and upper groups both show a double-peak enhancement feature. A detailed analysis of the nature and dynamical behavior of these QESs by Chu and Chu[18] reveals that the ionization enhancement is mainly due to the effect of the *charge-resonance-enhanced* multiphoton resonances of the $1\sigma_g$ and $1\sigma_u$ states with the excited electronic states at some particular inter-nuclear distances.

To evaluate individual contributions from the lower and upper sets of QESs to the overall ionization rates, in Figure 3, we have compared the overall ionization rates obtained in the present work (Figure 1) with the isolated ionization rates of these two QESs reported by Chu and Chu.[18] This comparison show that the lower QES dominantly determines the overall ionization rates for all of the $R < 3.6$ and $R > 5.2$ ranges. While, for the intermediate region $3.6 < R < 5.2$, both lower and upper QESs contribute competitively to the overall ionization rates. Therefore, the first peak of the R-dependent overall ionization rate curve observed in our calculations can be assigned to the upper QES, and the second and third peaks of the R-dependent overall ionization rate can be attributed to the lower QES.

Based on the instantaneous ionization rates calculated in this work, details of the time-dependent behavior of the system following the variations of the laser field can be extracted and used in the interpretation of the enhanced ionization rates. As can be seen from Figure 2, the enhanced ionization rate at about R=9.6 in Figure 1 can be attributed mainly to the several strong and sharp time-dependent ionization peaks. Existence of these strong sharp peaks is a direct evidence for the effect of charge-resonance-enhanced multiphoton resonances of the $1\sigma_g$ and $1\sigma_u$ states with the excited electronic states at some particular inter-nuclear distances.[14,18] These sharp peaks are not necessarily located at the maxima of the laser field oscillations. The baseline peaks at R=9.6, Figure 2(c), have large amplitudes without effective falling to near zero. While at other inter-nuclear distances, baseline peaks of the time-dependent ionization rates fall off effectively to very small values after each rising period (see Figure 2). As compared to R=9.6, amplitudes of the baseline peaks at other inter-nuclear separations, for example at R=6.0, are relatively small in spite of having some sharp peaks. Therefore, it can be concluded that the overall ionization rate increases as the number of strong sharp peaks and the level of the baseline of the time-dependent ionization rates increase.

On the basis of Eq. (15), positive (negative) value of instantaneous imaginary energy results in negative (positive) value for the instantaneous ionization rate, and therefore, the positive (negative) value of the instantaneous imaginary energy corresponds to the incoming (outgoing) of the electron to (from) the system. Designing, however, a simulation to present both negative and positive instantaneous ionization is a difficult task, because the outgoing and incoming electron wavepackets occur simultaneously and the intensity of the positive instantaneous ionization rates often overcomes the negative instantaneous ionization rates. Thus, in this section, we introduce a simulation to present both the negative and positive instantaneous ionization rates explicitly. These calculations of the (instantaneous) ionization rates was applied for boxes presented in Figures 4, i.e. with (640,170) size for the ($z, \rho$) directions. We now repeat simulation for R=9.6 in the linearly polarized laser field of I=$1.0 \times 10^{14}$ W/cm$^2$ and $\lambda$ = 1064 nm which is turned on immediately as $E_0 \cos(\omega t)$, i.e. with $f(t) = 1$. The simulation box is set to span over $-300 \leq z \leq 300$ and $0 \leq \rho \leq 150$, with the absorber potential regions set over a 20 au width strip around the borders, i.e. $-320 \leq z \leq -300$ and $300 \leq z \leq 320$ for the $z$-borders and $150 \leq \rho \leq 170$ for the $\rho$-border.

Calculation of the instantaneous ionization rate (IIR) can be calculated for any desired limited region (sub-boxes) inside the simulation box. In the present work, we have examined a variety of sub-boxes, introduced in Figure 4 (highlighted in gray, bordered in red), for the calculation of the IIR.

These simulation sub-boxes cover regions with $-50 \leq z \leq 50$ (a), $-300 \leq z \leq 50$ (b), $-300 \leq z \leq 300$ (c) and $-250 \leq z \leq 250$ (d) all with $0 \leq \rho \leq 150$, and $-300 \leq z \leq 300$ with $0 \leq \rho \leq 100$ (e); demonstrated respectively in Figures 4(a) to 4(e). In Figure 4, the hatched regions show the absorber regions and the two red points near the origin represent positions of the two H nuclei, having a 9.6 distance in this case.

The instantaneous energy of the system, for example for the sub-box (b), Figure 4(b), is calculated using Eq. (14) as

$$E(t) = \int_{\rho=0}^{\rho=150} \int_{z=-300}^{z=50} \psi(z,\rho,t) H(z,\rho,t) \psi^*(z,\rho,t) \rho d\rho dz . \qquad (17)$$

which has two real and imaginary components. The IIR is then calculated based on the imaginary part of the instantaneous energy using Eq. (15). The calculated IIR for the sub-boxes introduced in Figure 4 are plotted in Figure 5 over the time period of the first five cycles of the laser pulse.

We now analyze, as an example, the IIR plotted in Figure 5(a) corresponding to the sub-box (a) of Figure 4. When the laser field is turned on, the electron wavepacket starts outgoing from the borders of the sub-box (the gray region), i.e. passing from $z$ borders at -50 and 50 and the $\rho$ border placed at 150. Based on the sizes of this sub-box, we can expect that the wavepacket passes though the $z$ borders much earlier than that through the $\rho$ border. Note that the initial wavepacket in the absence of the laser field decays effectively to zero at the borders. For each coordinate, closer borders to the nuclei, or positions of the maximum probability of the wavepacket, result in sooner passage of the wavepacket from the borders. This, consequently, increases the outgoing part of the wavepacket and the ionization rate from the borders. Therefore, for the case of Figure 4(a), the outgoing part of the electron wavepacket from the $z$ borders should be larger than that from the $\rho$ border. For each cycle of the laser pulse, outgoing of the wavepacket occur periodically in two phases passing respectively through the two $z$ borders located at the $+z$ and $-z$ directions. Figure 5(a) shows a baseline peak of the IIR for every half cycle of the laser field. The intensity of these peaks reaches its asymptotic value during the first two cycles of the laser pulse. The first peak corresponding to the first half cycle of the laser pulse is relatively weak.

For the sub-box (b), introduced in Figure 4(b), the borders in the $+z$ and $-z$ directions are set respectively to +50 and –300. This setup of the z borders results in two effects. First, the ionization from the $-z$ border decreases considerable during the first few half cycles of the laser due to the longer time needed for the wavepacket to reach this border (ionization form the $+z$ border at 50 remains however the same as that for the sub-box (a)). The other effect is that the closer the sub-box borders to the concentrated part(s) of the initial wavepacket, the higher the intensity of the IIR peaks. Therefore, the contribution of the (negative or positive) ionization from the $+z$ border at +50 to the IIR is considerably stronger than that from the $-z$ border at –300. The overall result of these two effects is that, compared to the sub-box (a), the IIR peaks obtained for the sub-box (b) are alternatively intensified and weakened (or even become negative); this can be seen clearly from Figure 5(b). In other words, in the first few cycles of the laser pulse, no effective ionization occurs via the $-z$ border when the electric field vector pointing the $-z$ direction. In fact, during these half-cycles, the wavepacket returns back into the sub-box after passing through the $+z$ border of the sub-box and before being evolved over the outer region of the sub-box far from this border in the simulation box, while no considerable ionization (outgoing) of the wavepacket occur via the $-z$ border. The net result of these two counter-current wavepackets is thus a relative increase in the norm of the wavepacket inside the sub-box. This is equivalent to a negative ionization rate corresponding, via Eq. (15), to a positive value for the imaginary part of the energy of the system inside the sub-box, according to Eq.

(17). We can thus expect to have even negative ionization peaks for sub-box (b) corresponding to every other half-cycles of the laser pulse as can be seen evidently in Figure 5(b).

In the same time window of the first few cycles of the laser pulse, ionization of the wavepacket via the $+z$ border at +50 occurs normally as in (a) during the other set of the half-cycles, i.e. when the electric field vector pointing the $+z$ direction. Therefore, two sets of peaks should appear in the time trend of the IIR for this sub-box design which are located alternatively in time. Figure 5(b) exhibits that the positive IIR signals (corresponding to the dominant outgoing of electron wavepacket from the $+z$ border and the weak incoming of the electron wavepacket from the $-z$ border) are strong and sharp, but the negative IIR signals (corresponding to the incoming electron wavepackets from the $+z$ border and the outgoing electron wavepacket from the $-z$ border) are weak and smooth. Results obtained for this design of the sub-box inside the simulation box firmly justify existence of the negative IIR peaks.

If the borders of the sub-box are placed at sufficiently large distances from the nuclei, as in Figure 4(c), probability of having a return current of electron, after the wavepacket has passed through the borders and evolved over the outer part of the sub-box in the absorber regions, and consequently the probability of having negative IIR peaks vanishes, Figure 5(c). Farther borders result in smaller ionization in the early stages of the laser pulse and thus decrease the intensities of both positive and negative signals of the IIR. This can be deduced by comparing Figures 5(b) and 5(c). This comparison shows that the negative IIR peaks are effectively disappeared and oscillations of the positive signal are decreased. Analysis of Figures 5(a) and 5(c) shows that for the symmetric sub-boxes similar to sub-boxes (a) and (c) in Figure 4, during the every other half-cycles in which the electric laser field is in the $+z(-z)$ direction, the outgoing electron from the $+z$ $(-z)$ border dominates the incoming electron from the $-z$ $(+z)$ border resulting always in net positive IIR signals. This explains why observation of the negative IIR peaks is difficult in symmetric sub-boxes similar to Figures 5(a) and 5(c).

To accomplish our study on the effects of the region boundaries on the IIR signal, we set two other designs for the sub-box which are demonstrated in Figures 4(d) and 4(e). Results of our calculations of IIR for these two types of sub-boxes are presented in Figures 5(d) and 5(e), respectively. Comparison between Figures 5(c) and 5(d) shows that the absorber boundaries, the hatched regions, have negligible effect on the IIR signal. This can also be observed by comparing Figures 5(c) and 5(e). These comparisons approve the point that relative positions of the sub-box borders with respect to the concentrated part(s) of the initial wavepacket determine the IIR signal. Therefore, differences among Figures 5(c), 5(d) and 5(e) can be interpreted based on this relative positions only; the closer the sub-box borders to the concentrated part(s) of the initial wavepacket, the higher the intensity of the IIR peaks.

We can conclude from the results presented in Figures 4 and 5 that the IIR can be calculated for any desired limited sub-box inside the simulation box, and the instantaneous behavior of the electron wavepacket in intense laser field can be probed closely using the IIR of the system, in the space-time.[29]

## Conclusion

We introduced and calculated instantaneous ionization rate based on the instantaneous values of the imaginary energy. It is shown that positive and negative instantaneous imaginary energies of the system represent respectively the incoming (returning) and the outgoing instantaneous currents of electron. Thus, the instantaneous behavior of the electron wavepacket in intense laser field can be probed closely using the instantaneous imaginary energy of the system. In this approach, one can determine the instantaneous intensity and direction of the electronic current. We can also determine the delay time of the outgoing electron wavepackets with respect to the corresponding peaks of the laser field. The instantaneous ionization rate is used to study details of the enhanced ionization rates observed at specific inter-nuclear distances. This approach gives direct evidence for the existence of the effect of charge-resonance-enhanced multiphoton resonances of the QESs with excited electronic states at some particular inter-nuclear distances. TDSE and ac Floquet calculations together form a powerful tool to derive detailed mechanism of the ionization enhancement phenomenon.[29] The comparison between the results obtained by these two methods lead us to determine the individual contributions of the QESs to the overall ionization rates.

## Acknowledgments

We would like to acknowledge the University of Isfahan for providing financial supports and research facilities. We should also acknowledge Scientific Computing Center of the School of Mathematics, Institute for Studies in Theoretical Physics and Mathematics (IPM) of I. R. Iran for providing High Performance Computing Cluster to carry out calculations.

## Figure Captions:

**Figure 1.** The ionization rate, $\Gamma$, of $H_2^+$ as a function of R in the linearly polarized field of $I = 1.0 \times 10^{14}$ W/cm$^2$ intensity and $\lambda = 1064$ nm wavelength calculated in this work (●) compared with the calculated ionization rates reported by Lu and Bandrauk ($\Delta$) [15], Peng et al. (□) [16] and Chu and Chu (×) [18]. The isolated single point on the right vertical axis (at R=14) corresponds to the ionization rate of an isolated H atom.

**Figure 2.** The instantaneous ionization rates of $H_2^+$ in the linearly polarized laser field of $I = 1.0 \times 10^{14}$ W/cm$^2$ intensity and $\lambda = 1064$ nm wavelength over a two-cycle period, from 14.0 to16.0 cycles. Note the different scales used for different values of R.

**Figure 3.** A comparison between the overall ionization rates obtained in this work (●) and individual contributions of the upper and lower quasi-energy states (QESs) reported by Chu and Chu [18].

**Figure 4.** The different sub-boxes (highlighted in grey and bordered in red) designed for probing the evolution and ionization of the wavepacket of the electron in a simulation box of $-300 \leq z \leq 300$ and $0 \leq \rho \leq 150$. The hatched strips show the absorber regions. The two red points near the origin represent the two H nuclei located with a 9.6 distance in this case. These boxes span over (a) $-50 \leq z \leq 50$, (b) $-300 \leq z \leq 50$, (c) $-300 \leq z \leq 300$ and (d) $-250 \leq z \leq 250$ all with $0 \leq \rho \leq 150$, and over (e) $-300 \leq z \leq 300$ with $0 \leq \rho \leq 100$. The instantaneous ionization rates calculated for these sub-boxes are presented in Figure 5.

**Figure 5.** The instantaneous ionization rates corresponding to the sub-boxes introduced in Figure 4. The simulations are carried out for R=9.6 for the first five cycles of the linearly polarized laser pulse of I= $1.0 \times 10^{14}$ W/cm$^2$ intensity and $\lambda$ = 1064 nm wavelength which is turned on immediately as $E_0 \cos(\omega t)$, i.e. with $f(t) = 1$.

**Figure 1**

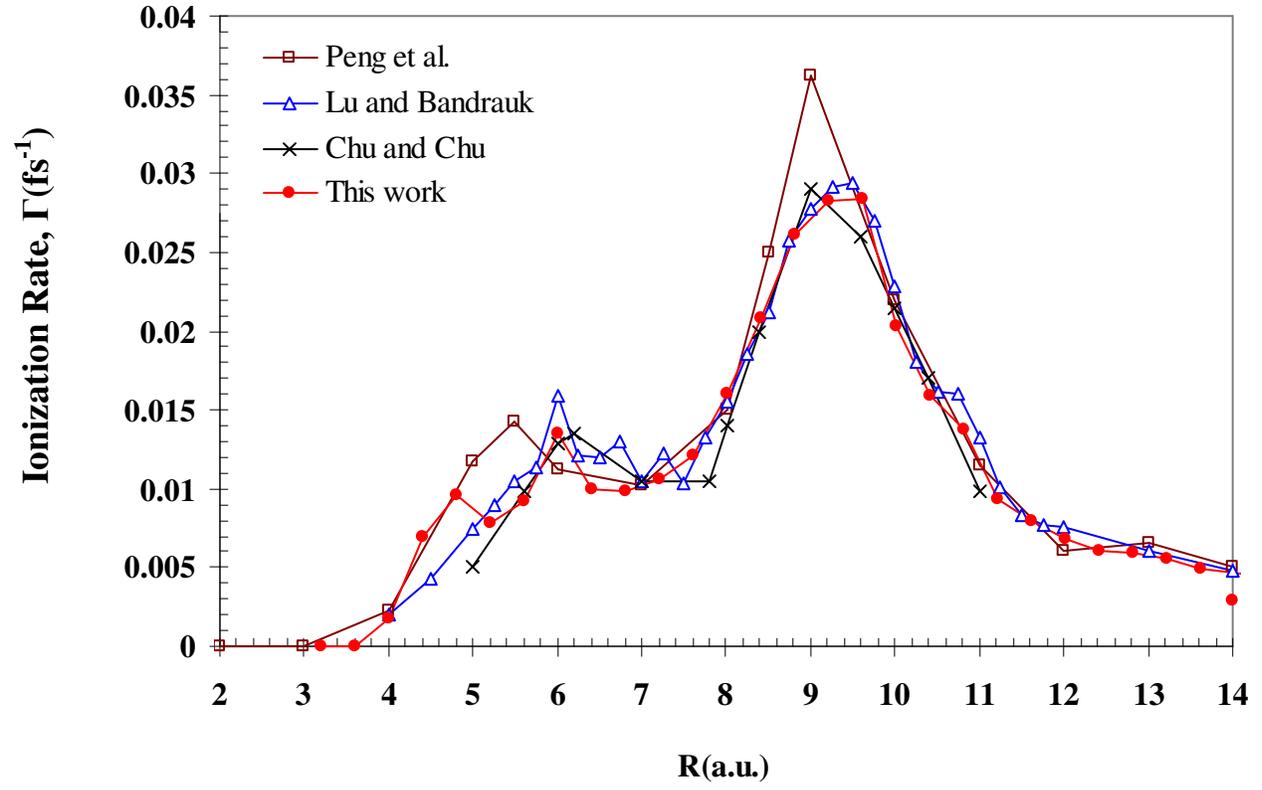

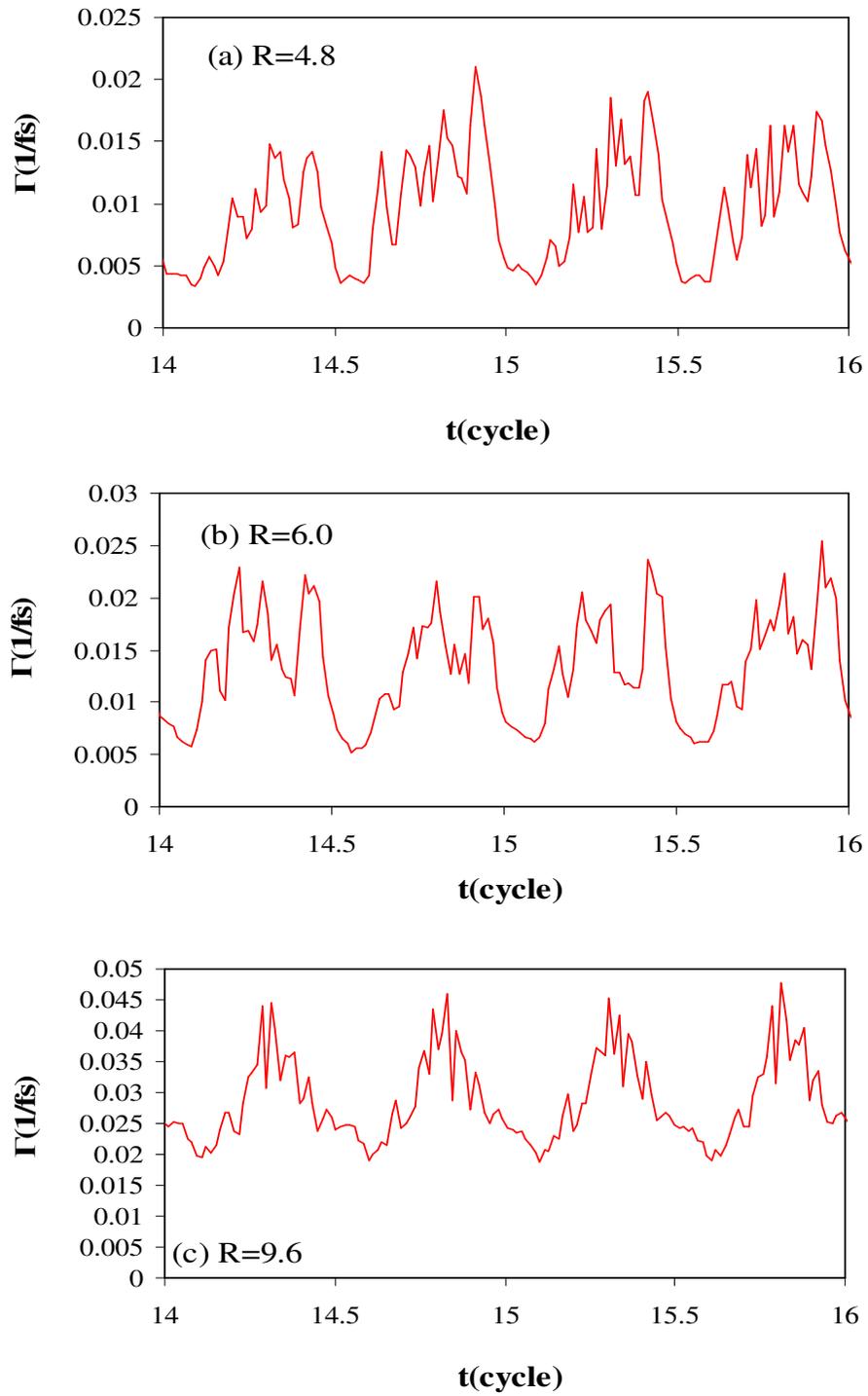

Figure 2

**Figure 3**

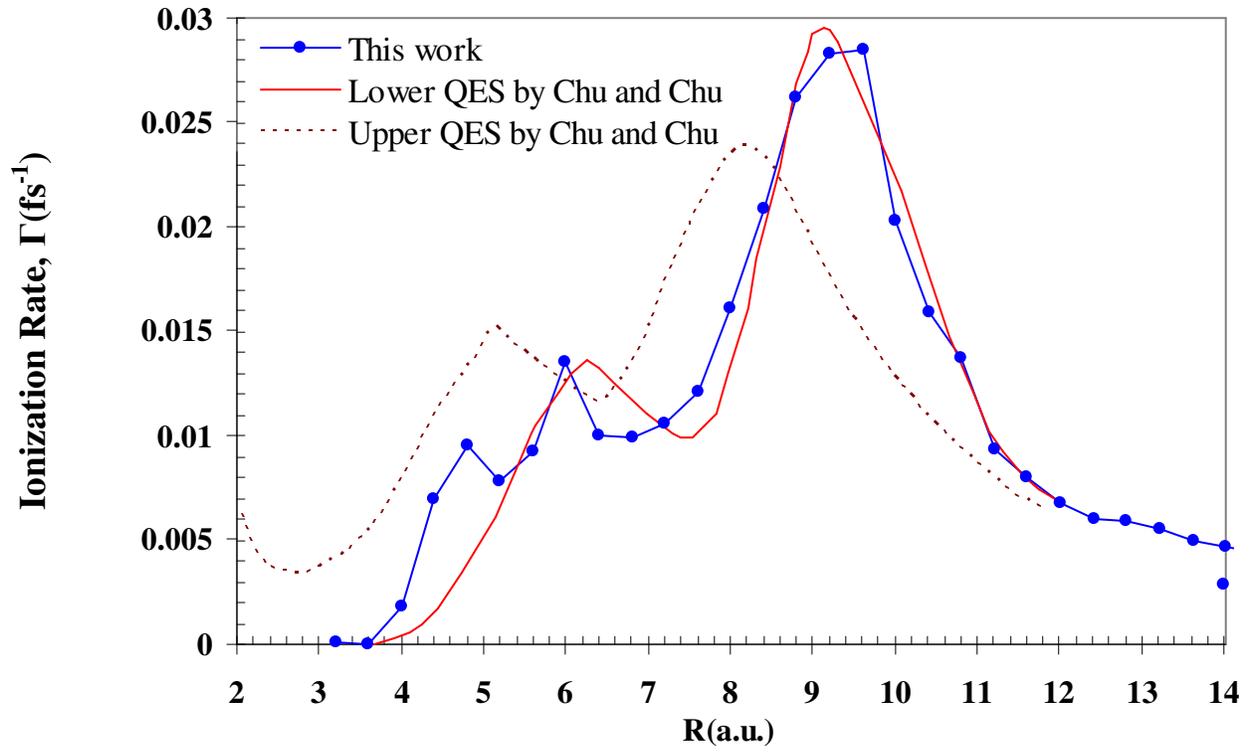

**Figure 4**

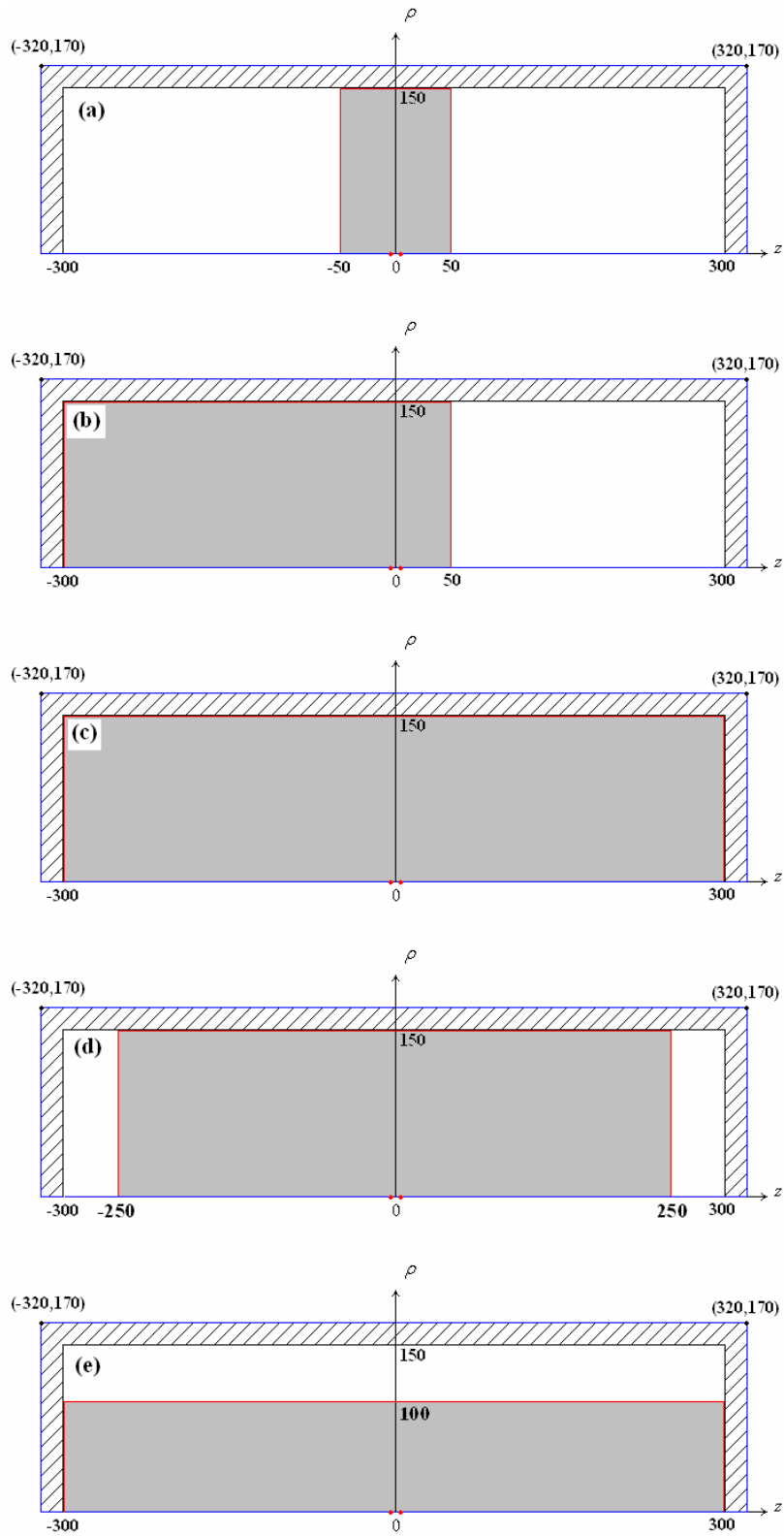

**Figure 5**

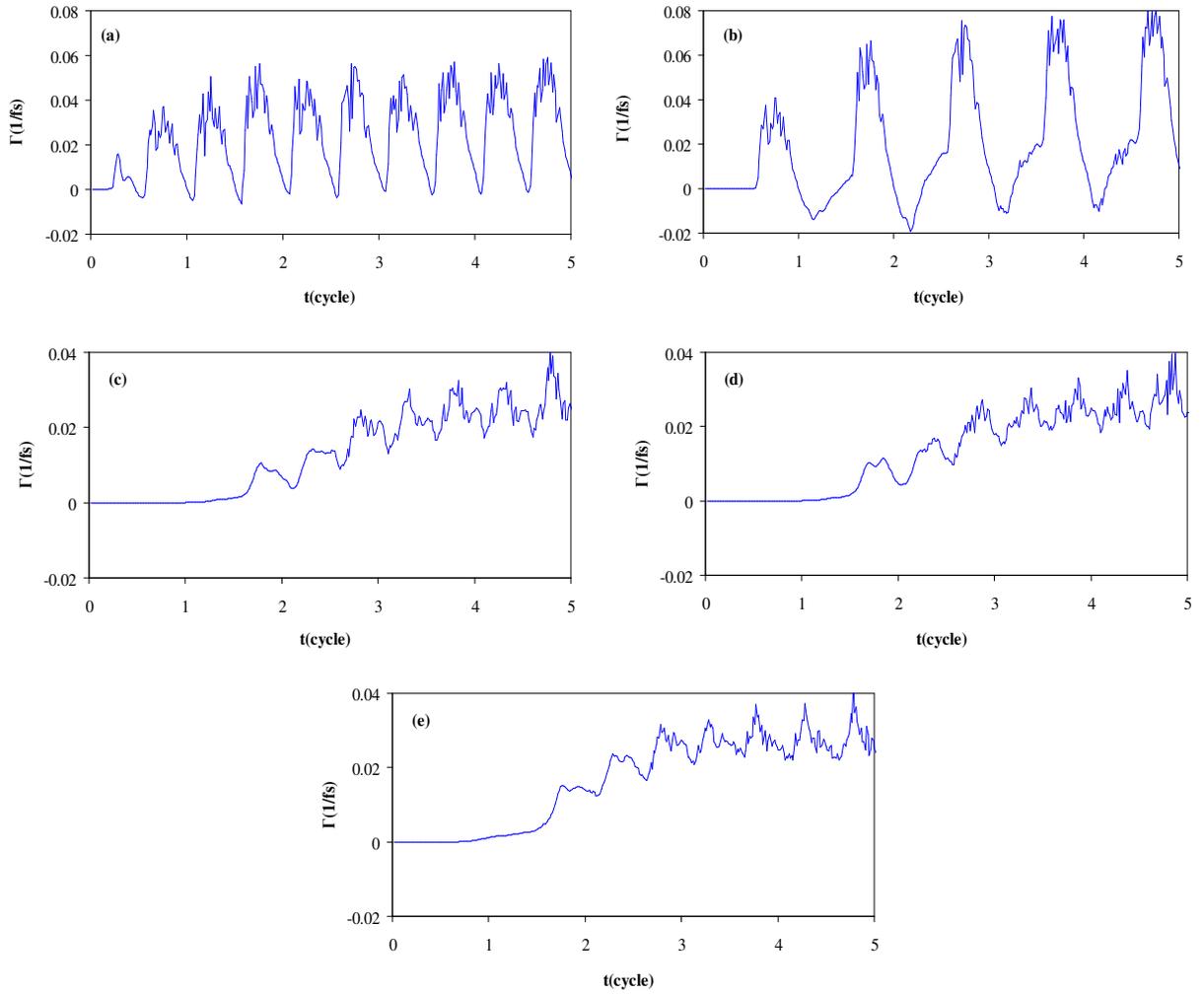